%% file: main.tex
\documentclass[11pt,a4paper]{vcbench}
\usepackage[numbers,sort&compress]{natbib}
\usepackage{graphicx,caption,float,booktabs,tabularx,longtable,array,ragged2e}
\usepackage{amsmath,xspace,microtype}
\usepackage{placeins}
\DeclareCaptionLabelFormat{continued}{#1~#2 \emph{(continued)}}

\title{VCBench: Benchmarking virtual cell models for in-the-wild perturbation response}
\correspondingauthor{
$\spadesuit$: Core Contributor \quad
$\clubsuit$: Corresponding Author \\ 
Please send correspondence regarding this report to sunsiqi1@pjlab.org.cn; gaozhangyang@pjlab.org.cn; shuizhou@shu.edu.cn
}

\author[1,2,3$\spadesuit$]{Xinjie Mao}
\author[1,4$\spadesuit$]{Songming Zhang}
\author[5]{Qianhong Wen}
\author[6]{Xiangyu Wen}
\author[7]{Kedu Jin}
\author[1,8]{Sheng Xu}
\author[1,8]{Hao Wu}
\author[1]{Lang Yu}
\author[2]{Yuqiang Li}
\author[1]{Lei Bai}
\author[9]{Qi Liu}
\author[1,10]{Ning Ding}
\author[1,8 $\clubsuit$]{Siqi Sun}
\author[1 $\clubsuit$]{Zhangyang Gao}
\author[1,11 $\clubsuit$]{Shuizhou Chen}

\affil[1]{Shanghai Artificial Intelligence Laboratory}
\affil[2]{Shanghai Innovation Institute}
\affil[3]{School of Life Sciences, Westlake University}
\affil[4]{Faculty of Computer Science and Control Engineering, Shenzhen University of Advanced Technology, Shenzhen}
\affil[5]{College of Information Engineering, Sichuan Agricultural University}
\affil[6]{School of Electronics and Information Technology, Sun Yat-sen University}
\affil[7]{School of Life Science and Technology, China Pharmaceutical University}
\affil[8]{Research institute of intelligent complex systems, Fudan University}
\affil[9]{Shanghai Skin Disease Hospital, School of Medicine \& Frontier Science Center for Stem Cell Research, Bioinformatics Department, School of Life Sciences and Technology, Tongji University}
\affil[10]{Department of Electronic Engineering, Tsinghua University}
\affil[11]{School of Computer Engineering \& Science, Shanghai University}

\newcolumntype{Y}{>{\RaggedRight\arraybackslash}X}
\hypersetup{colorlinks=true,linkcolor=MidnightBlue,citecolor=MidnightBlue,urlcolor=MidnightBlue,pdftitle={VCBench: Benchmarking virtual cell models for in-the-wild perturbation response},pdfauthor={Mao et al.}}
\fancypagestyle{firststyle}{\fancyhf{}\fancyfoot[R]{\thepage}}
\begin{abstract}
\input{Section/0_abstract}
\end{abstract}
\begin{document}
\maketitle
\par\smallskip{\footnotesize\the\correspondingauthor\par}\medskip
\input{Section/1_Introduction}
\input{Section/2_results}
\input{Section/3_discussion}
\input{Section/4_methods}
\bibliographystyle{unsrtnat}
\bibliography{main}
\clearpage
\input{Section/X_appendix}
\end{document}

%% file: Section/0_abstract.tex
Virtual cell (VC) models aim to predict cellular responses to diverse perturbations in silico, offering great promise for drug discovery and precision medicine. However, although models routinely report strong performance on existing benchmarks, whether these predictions translate to practical biological utility remains unclear. Current evaluations often rely on overly simplified or inconsistent setups that fail to capture the complexity and variability of real biological systems. Here, we present a standardized, modular benchmarking framework that evaluates nine VC models under challenging, in-the-wild scenarios—spanning unseen cell lines and donors, novel genetic and chemical perturbations, and cross-study transfer. Across these evaluations, complex models do not consistently surpass simple linear baselines on broad transcriptional trends, while fine-grained, perturbation-specific effects remain difficult to recover. Moreover, different evaluation metrics emphasize distinct biological properties, resulting in markedly different model rankings. Consequently, model selection and data integration strategies should be tailored to specific downstream tasks rather than applied uniformly. Together, this framework establishes a more biologically grounded benchmark to guide model development, evaluation, and deployment in real-world settings.

%% file: Section/1_Introduction.tex
% \phantomsection
\section{Introduction}
\label{sec:introduction}

Virtual cell models aim to predict cellular responses to genetic and chemical perturbations in silico, supporting efforts to map gene function and understand drug response~\cite{rood2024toward}. Their development builds on single-cell perturbation screens—such as CRISPR-based Perturb-seq~\cite{dixit2016perturb,adamson2016multiplexed,replogle2022mapping} and chemical profiling platforms~\cite{srivatsan2020massively}—that measure transcriptional changes across many conditions at single-cell resolution. Even large screens, however, sample only a small fraction of all possible perturbation–context combinations. This combinatorial sparsity motivates computational models that generalize from observed measurements to predict unseen conditions and prioritize follow-up experiments.

Predicting responses under unseen conditions remains difficult because cellular states shape how perturbations act. The same perturbation can elicit distinct effects across cell types or genetic backgrounds, and responses to multi-gene perturbations often deviate from additive combinations of single-perturbation effects~\cite{rood2024toward}. Moving across studies introduces further discrepancies in experimental protocols, profiled genes, and perturbation sets, altering both the training and the prediction targets. Together, these factors complicate assessing how reliably models generalize beyond training conditions, and, by extension, determining when and how combining data across contexts improves prediction.

Recent benchmarks have evaluated perturbation models across diverse datasets and generalization splits using complementary metrics~\cite{wei2025benchmarking,wu2025perturbench}, while also exposing subtleties in how scores are interpreted. Simple linear models and mean-expression baselines can match or outperform more complex architectures in certain settings~\cite{ahlmanneltze2025deep,wong2025simple}, whereas metrics tailored to perturbation-specific responses can uncover advantages of deep learning models~\cite{miller2026calibrated}. Yet existing benchmarks still leave open how representation-distance shifts, fine-grained response magnitude errors, and controlled cross-study integration shape model behavior. Consequently, a high aggregate score does not by itself confirm that a model identifies affected genes or recovers the direction and magnitude of their changes. Score interpretation also depends on how closely test conditions align with the training data—whether a held-out cell line resembles seen cellular contexts or an unseen perturbation shares features with trained perturbations. Rigorous evaluation therefore requires characterizing condition shifts alongside the specific response properties captured by each model.

To address these gaps, we introduce VCBench, a modular benchmarking framework designed to diagnose how virtual cell models generalize and how evaluation choices shape model conclusions. Built around standardized data, model, and evaluation interfaces, VCBench enables consistent comparisons across diverse architectures under realistic distribution shifts—spanning unseen cellular backgrounds, novel genetic and chemical perturbations, and cross-study transfer. Rather than relying on a single aggregate score or random splits alone, the framework pairs embedding-guided context holdouts with cross-study integration experiments and evaluates complementary response properties, including global transcriptional agreement, affected-gene recovery, and the direction and magnitude of expression changes. By systematically examining how split design, training-source selection, and metric choice alter observed performance, VCBench provides a practical lens for interpreting model capabilities and guiding data and model choices for specific biological tasks.

% Our evaluation shows that predictive performance depends on both the conditions being tested and the response properties being measured. In the evaluated cell-context tests, embedding-guided holdouts reveal poorer recovery of affected genes than random holdouts, exposing limitations in generalization across cellular backgrounds. Overall agreement between predicted and measured responses can coexist with errors in identifying affected genes or estimating the sizes of their changes. Cross-study comparisons further show that combining datasets does not consistently improve prediction, with outcomes varying across data sources and integration strategies. Model rankings also differ across evaluation measures. Together, these findings support characterizing the differences between training and test conditions, evaluating complementary aspects of the response, and choosing training data and models for a defined prediction task.

We find that predictive performance varies with both the tested conditions and the measured response properties. In cell-context evaluations, embedding-guided holdouts show poorer recovery of affected genes than random holdouts, highlighting limited generalization across cellular backgrounds. Meanwhile, even simple linear approaches can capture global transcriptional trends, yielding strong overall agreement that nonetheless coexists with substantial errors in identifying affected genes or estimating fine-grained, perturbation-specific effect sizes. Cross-study analyses further reveal that combining datasets yields inconsistent gains depending on data sources and integration strategies, and that model rankings depend heavily on the choice of evaluation metric. Consequently, robust benchmarking requires accounting for distribution shifts between training and test conditions, evaluating complementary response properties, and selecting data and models tailored to the specific prediction task.

%% file: Section/2_results.tex
% \clearpage
\section{Results}
\label{sec:results}

\subsection{Benchmark design for in-the-wild virtual cell prediction}
\label{sec:framework}

\begin{figure}[!htbp]
    \centering
    \includegraphics[width=.9\linewidth]{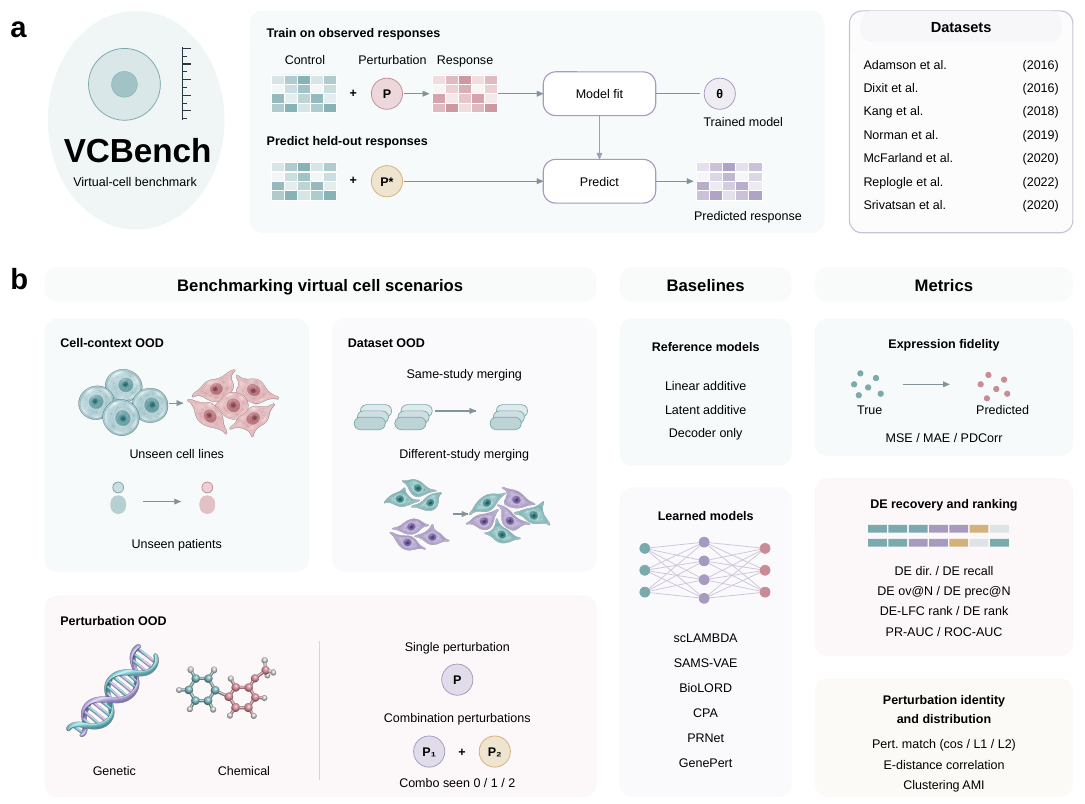}
    \caption{\textbf{a,} Training on observed control and perturbed cells and prediction of held-out perturbation responses. \textbf{b,} Cell-context, genetic perturbation, chemical perturbation and cross-study prediction settings.
    }
    \label{fig:workflow}
\end{figure}

We developed VCBench to evaluate how accurately virtual cell models predict perturbation responses beyond the conditions used for training. The framework brings model training, prediction and evaluation into a common workflow: models are trained on observed control and perturbed cells, and their predictions are evaluated against measured responses withheld from training (Fig.~\ref{fig:workflow}a).

VCBench organizes out-of-distribution (OOD) evaluation around three broad generalization settings (Fig.~\ref{fig:workflow}b). Cell-context tests assess prediction in held-out cell lines or donors. At the perturbation level, genetic tests withhold single-gene perturbations or gene combinations, while chemical tests withhold compounds. Cross-study tests use training data from multiple studies to predict responses in a separate target study. 

Within cell-context tests, an unseen cell line or donor may still resemble backgrounds represented in training. To characterize this similarity, we represent each cell line or donor using embeddings derived from its control-cell expression profiles. These control cells provide the reference expression profiles for the corresponding perturbation experiment. Using these representations, we compare randomly selected holdouts with embedding-guided partitions that select test backgrounds with greater separation from training backgrounds in the chosen embedding space. Training, validation and test sizes are matched between the two split strategies.

Across these tasks, the core benchmark includes nine models, comprising six perturbation predictors and three reference baselines, evaluated across nine data packages from seven studies (Tables~\ref{tab:models} and~\ref{tab:datasets}). Within each task, the evaluated models use the same expression features and perturbation annotations and are scored against the same held-out responses. The three reference baselines provide comparison points for assessing whether the other models offer additional predictive value.

To assess both overall agreement and the recovery of specific perturbation effects, we combine measures of expression error, response correlation, differential expression (DE) and perturbation identification. Expression-error metrics quantify differences between predicted and observed expression, while response correlation measures agreement in expression changes relative to matched controls. DE metrics assess recovery of affected genes, agreement in fold-change direction and concordance of fold-change rankings. We additionally assess whether a prediction can be matched to the correct observed perturbation response among candidate responses. These cell-population and DE metrics are computed using \texttt{cell\_eval} v0.6.6~\cite{adduri2026state}, with formal definitions provided in the Methods and Supplementary Table~\ref{tab:metriccatalog}.

% \FloatBarrier

\subsection{Models recover fewer affected genes in more distinct cell backgrounds}

Models recovered fewer affected genes on average under embedding-guided holdouts of cell lines and donors, even when overall response correlation changed little (Fig.~\ref{fig:context}c). We examined this contrast by comparing random and embedding-guided holdouts in McFarland cancer cell lines and Kang donors (Fig.~\ref{fig:context}a).

\begin{figure}[p]
  \centering
  \includegraphics[width=\linewidth,height=0.95\textheight,keepaspectratio]{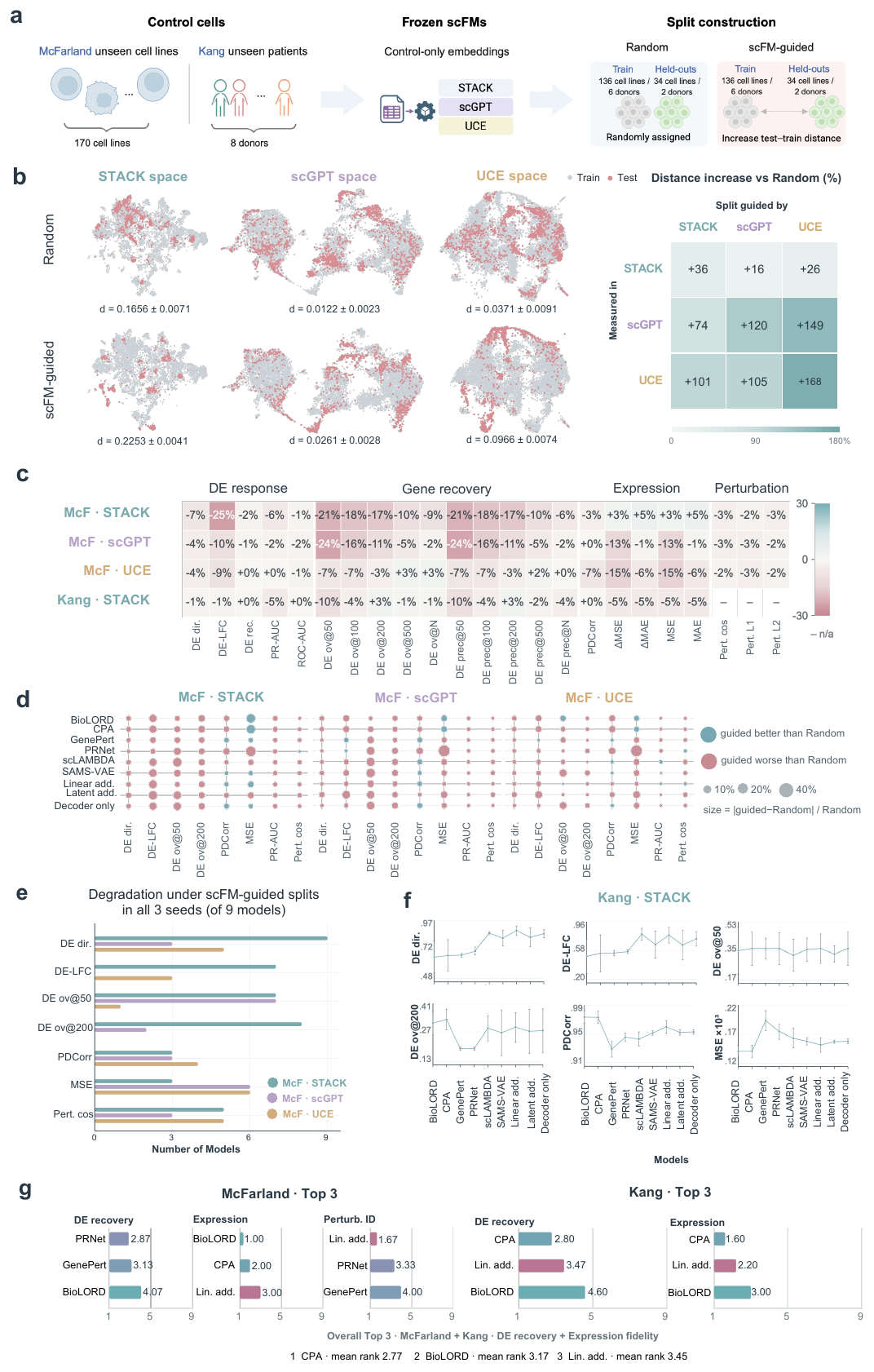}
  \caption{\textbf{Prediction in unseen cell lines and donors.}}
  \label{fig:context}
\end{figure}

\begin{figure}[!t]
  \ContinuedFloat
  \captionsetup{labelformat=continued}
  \caption{\textbf{a,} Random and control-representation-guided holdouts at matched partition budgets.
\textbf{b,} UMAP examples and nearest-training cosine-distance audits; the matrix shows relative distance increases over Random. Distance annotations are mean $\pm$ SD across three seeds.
\textbf{c,} Direction-adjusted relative performance changes averaged over nine models and three seeds; dashes indicate unavailable metrics since Kang only has 1 perturbation.
\textbf{d,} Model-level changes; bubble area encodes absolute relative change, with teal for improvement and rose for deterioration.
\textbf{e,} Counts of models with deterioration in all three seed comparisons, out of nine. For McFarland STACK, the counts are nine for DE direction and eight for DE ov@200.
\textbf{f,} Kang STACK-guided raw scores, mean $\pm$ seed SD; lower MSE is better.
\textbf{g,} Top-three rankings by metric family and the combined McFarland/Kang summary; lower rank is better. Relative changes are expressed as percentages of the absolute Random score.}
\end{figure}

Both strategies withheld entire cell lines or donors. The difference was how these identities were chosen: a randomly held-out cell line can still closely resemble a training cell line. To select less similar backgrounds while avoiding leakage from held-out perturbation targets, we used embeddings derived from control-cell expression to capture baseline cellular state differences. We chose STACK~\cite{dong2026stack}, scGPT~\cite{cui2024scgpt} and UCE~\cite{rosen2026uce} for McFarland, and used STACK for Kang; training, validation and test sizes were matched between split strategies to ensure a controlled comparison (Fig.~\ref{fig:context}a). In McFarland, the average distance from held-out cell lines to their nearest training counterparts increased in all three embedding spaces, including spaces not used to construct a given split (Fig.~\ref{fig:context}b).

In the STACK-guided McFarland comparison, overlap between the predicted and observed top-50 DE gene sets was 21\% lower than under random holdouts, and fold-change rank agreement was 25\% lower. By comparison, perturbation-response correlation (PDCorr, which measures agreement between predicted and observed expression changes relative to matched controls) decreased by 3\% (Fig.~\ref{fig:context}c). These values are mean relative changes across nine models and three seeds. Kang showed a similar contrast: top-50 overlap was 10\% lower, while PDCorr changed little (Fig.~\ref{fig:context}c). The absolute scores under the guided Kang split further illustrated this distinction: all nine models had PDCorr values above 0.9, while top-50 DE overlap scores remained close to 0.35 (Fig.~\ref{fig:context}f).

The decrease in gene recovery also appeared across individual models (Fig.~\ref{fig:context}d). Under the STACK-guided McFarland split, top-50 overlap was lower in all three seed comparisons for seven of the nine models, and all nine predicted the direction of DE changes less accurately in every comparison (Fig.~\ref{fig:context}e). Mean squared error (MSE) nevertheless improved slightly on average in the same setting (Fig.~\ref{fig:context}c). Thus, lower expression error did not ensure better identification of affected genes.

The choice of metric also affected model rankings (Fig.~\ref{fig:context}g). PRNet ranked first for DE recovery in McFarland, whereas BioLORD ranked first for expression-level performance. Linear Additive ranked third in the combined McFarland--Kang ranking of DE recovery and expression fidelity, behind CPA and BioLORD. Its performance supports retaining an additive baseline when assessing the predictive value of more complex models.

Together, these tests show that models can retain high response correlation while becoming less accurate at identifying affected genes in unfamiliar cell backgrounds. Evaluating this capability requires specifying how test backgrounds differ from training data and checking which genes respond and whether their predicted changes have the correct direction.

% \FloatBarrier
% \phantomsection
\subsection{Limited generalization to unseen genetic and chemical perturbations}
\label{sec:perturbation}

We next tested whether models could predict gene-expression changes after a perturbation that was not included in training. We used single-gene and paired-gene perturbations in Norman K562 cells~\cite{norman2019exploring} and compound treatments in SciPlex3 A549, K562 and MCF7 cells~\cite{srivatsan2020massively} (Fig.~\ref{fig:perturbation}a,f). Training included cells from the test cell line, but excluded responses to the test perturbation. Each prediction covered expression changes across the evaluated genes, even when only one gene was perturbed.

\begin{figure}[p]
  \centering
  \includegraphics[width=\linewidth,height=0.95\textheight,keepaspectratio]{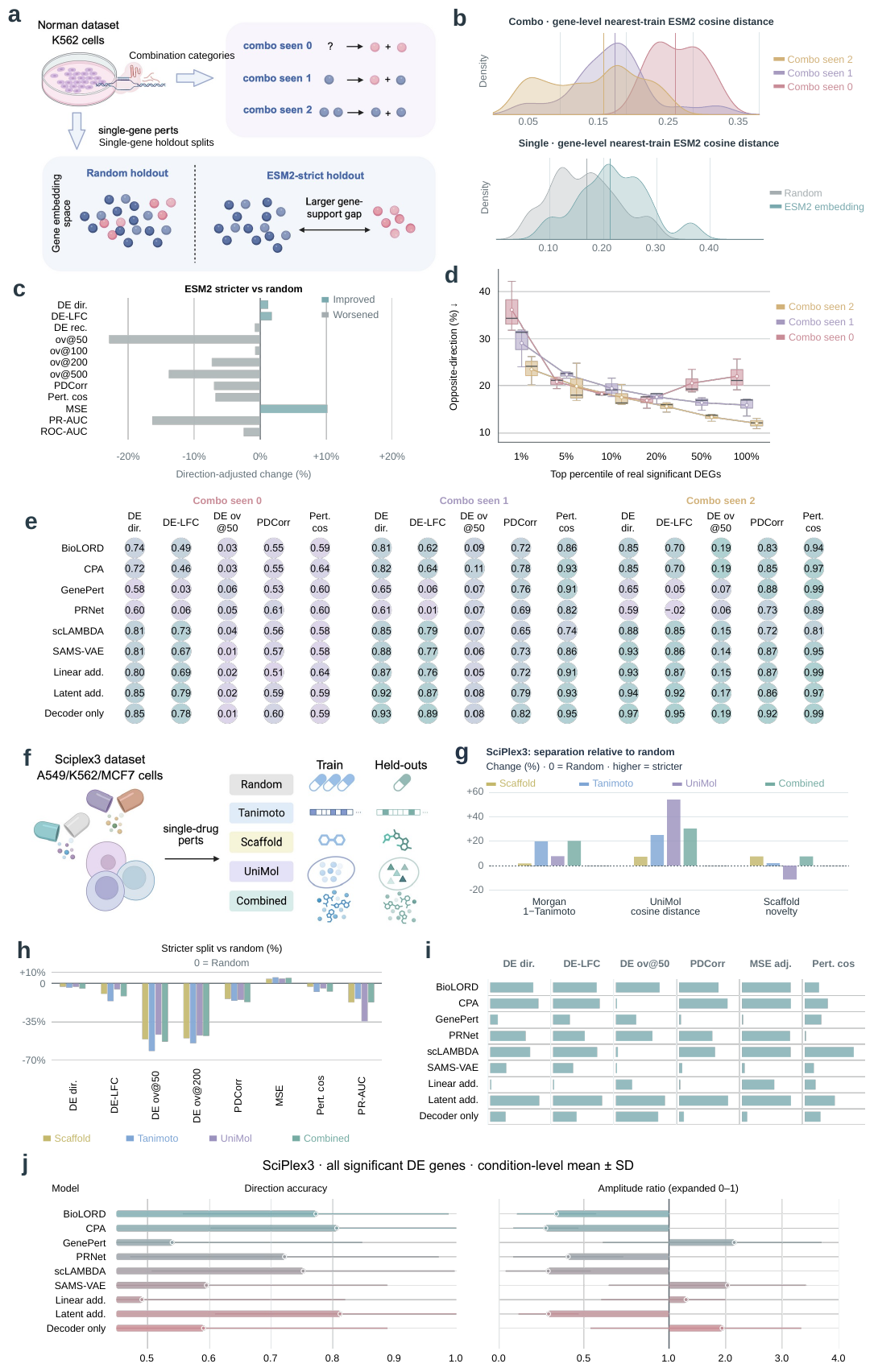}
  \caption{\textbf{Genetic and chemical perturbation holdouts.}}
  \label{fig:perturbation}
\end{figure}

\begin{figure}[!t]
  \ContinuedFloat
  \captionsetup{labelformat=continued}
  \caption{\textbf{a,} Norman single-gene and combination test designs. Seen 0, 1 and 2 count genes in a test pair that occur in training perturbations, either alone or in another pair; the test pair itself is withheld.
\textbf{b,} Gene-level nearest-training ESM2 cosine-distance densities for combination constituents and single-gene tests; panels use separate horizontal scales.
\textbf{c,} Direction-adjusted relative differences from Random for single genes, calculated from scores averaged across models and seeds.
\textbf{d,} Opposite-direction fractions by percentile of observed significant DE genes ranked by absolute observed fold change; boxes show seed quartiles and medians, whiskers the range, and connected points the means.
\textbf{e,} Comparison across models and gene-pair settings (means over three seeds). Each metric uses a shared color scale.
\textbf{f,g,} SciPlex3 split designs and separation relative to Random, including scaffold novelty.
\textbf{h,} Balanced-context relative performance changes; positive denotes improvement, including reduced MSE.
\textbf{i,} Combined model profiles, min--max normalized within each oriented metric.
\textbf{j,} Direction accuracy and amplitude ratio for observed significant DE genes with nonzero responses in the Combined split. The amplitude ratio is the predicted response $L_2$ norm divided by the observed response $L_2$ norm across these genes within each condition. Points and error bars show mean $\pm$ SD across conditions and seeds}
\end{figure}

\paragraph{Responses to single-gene perturbations}

For single-gene perturbations, we compared random holdouts with holdouts selected using ESM2 representations of the proteins encoded by the target genes. The latter increased the average distance from test targets to their nearest training targets in this representation, while preserving partition sizes (Fig.~\ref{fig:perturbation}a,b). Relative to Random, mean response MSE was approximately 10\% lower, but mean top-50 DE overlap was 23\% lower and response correlation (PDCorr) was 7\% lower (Fig.~\ref{fig:perturbation}c). These relative differences were calculated from scores averaged across models and seeds. Thus, a lower average expression error did not imply more accurate identification of affected genes.

\paragraph{Responses to gene combinations}

A gene combination can be new even when its individual genes have appeared in training perturbations. We therefore grouped held-out gene pairs according to whether zero, one or both constituent genes occurred in training, either alone or in another pair (Fig.~\ref{fig:perturbation}a). The specific test pair was absent from training in all three groups. Among the top 1\% of observed significant DE genes ranked by absolute fold change, mean direction-error rates were approximately 36\%, 29\% and 23\% for the zero-, one- and two-seen groups, respectively (Fig.~\ref{fig:perturbation}d). All nine models also had higher mean PDCorr in groups with more constituent genes represented in training (Fig.~\ref{fig:perturbation}e).

However, even when both genes were represented in training, mean top-50 DE overlap remained below 0.20 for every model (Fig.~\ref{fig:perturbation}e). The reference baselines remained competitive: Latent Additive and Decoder Only had among the highest direction and fold-change rank scores when neither gene was represented in training. These comparisons show that overall prediction scores vary with the information available about a combination, while identifying its affected genes remains difficult.

\paragraph{Responses to new compounds}
\label{sec:chemical_ood}

For compounds, a new identity can still share a chemical scaffold---the core ring systems and connecting linkers---with training molecules. We therefore compared random compound holdouts with four strategies based on molecular structure or molecular representations, using matched partition sizes (Fig.~\ref{fig:perturbation}f). The \emph{Combined} strategy kept compounds with identical fingerprints or shared nonempty scaffolds in the same partition and increased separation in both Morgan-fingerprint and UniMol representations. Consequently, no nonempty scaffold was shared between its training and test compounds, and average nearest-training distances were greater than under Random (Fig.~\ref{fig:perturbation}g).

Under \emph{Combined}, mean top-50 DE overlap and PDCorr were lower than under Random, despite a lower mean response MSE (Fig.~\ref{fig:perturbation}h). Model performance within this split also differed across metrics (Fig.~\ref{fig:perturbation}i). As in the single-gene comparison, a lower average error therefore did not ensure better identification of affected genes.

We further examined whether correctly predicting the direction of a drug response also meant predicting its size. On observed significant DE genes in the \emph{Combined} split, CPA and Latent Additive achieved mean direction accuracies of approximately 81\%, yet their mean amplitude ratios were below 0.30 (Fig.~\ref{fig:perturbation}j). This ratio compares the sizes of the predicted and observed response vectors, measured by their $L_2$ norms across these genes within each condition (Methods). Linear Additive had the mean ratio closest to one, but its mean direction accuracy was only approximately 49\%. Thus, neither correct directions nor a similar overall response size alone ensured accurate predictions for individual genes.

Together, the genetic tests show that better overall predictions in groups with more constituent genes represented in training can coexist with limited recovery of affected genes. The chemical tests further show that predicting the direction of a change correctly does not ensure that its size is estimated accurately.

% \FloatBarrier
% \phantomsection
\subsection{Aligning gene expression data improves cross-study prediction}
\label{sec:multisource}

To evaluate prediction across studies, we used four K562 datasets from Adamson~\cite{adamson2016multiplexed} and Dixit~\cite{dixit2016perturb} to predict perturbation responses in Replogle~\cite{replogle2022mapping}. Despite all profiling K562 cells, these datasets differed in the distribution of gene expression changes after perturbation (Fig.~\ref{fig:multisource}b). We compared two ways of preparing the training data: combining the datasets directly (Direct) or first adjusting their gene expression values using control cells as a reference. Each of the six prediction models was trained on the same cells and evaluated on the same Replogle test experiments under both approaches.

\begin{figure}[p]
  \centering
  \includegraphics[width=\linewidth,height=0.95\textheight,keepaspectratio]{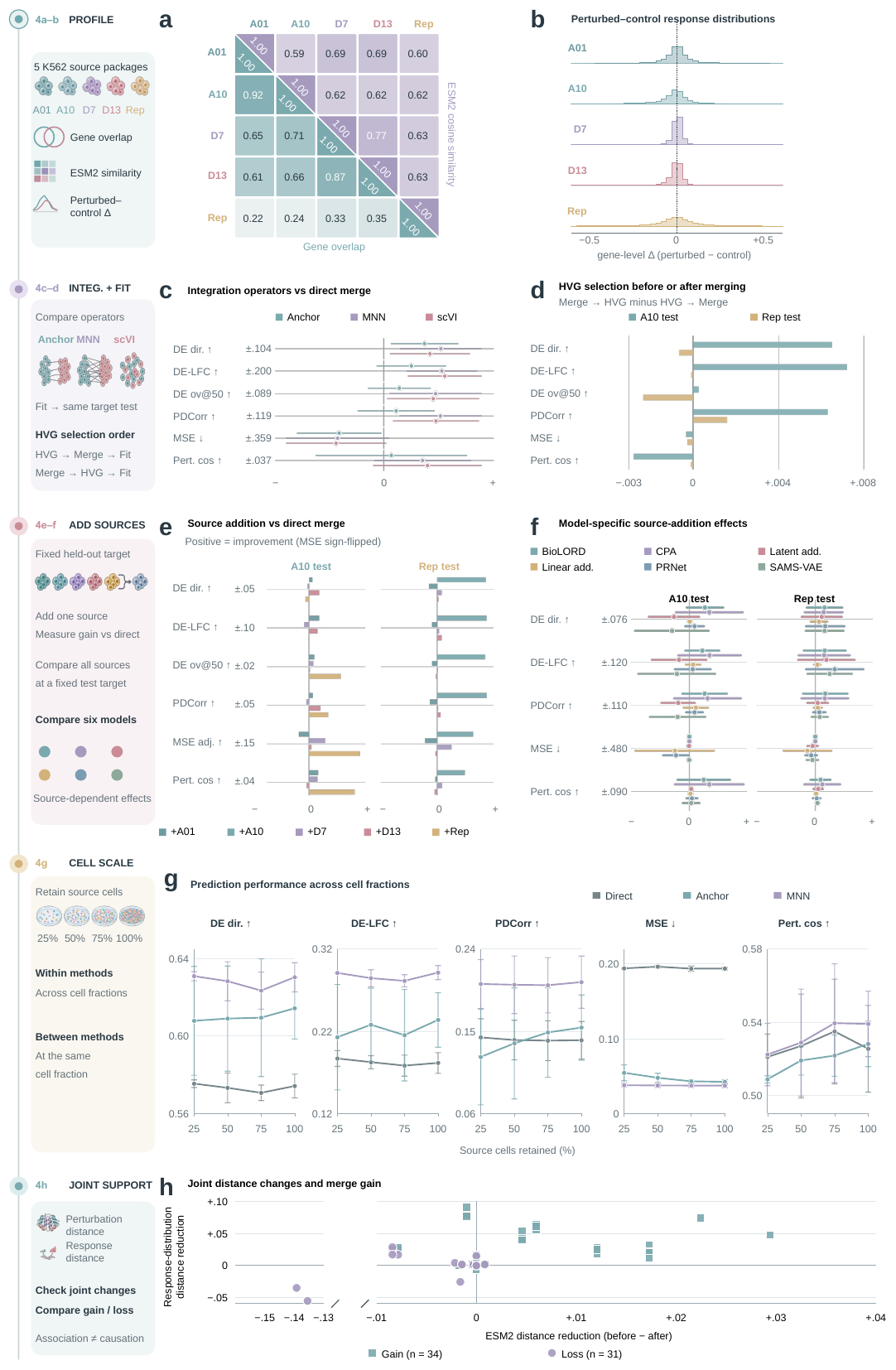}
  \caption{\textbf{Cross-study differences and source merging in K562.}}
  \label{fig:multisource}
\end{figure}

\begin{figure}[!t]
  \ContinuedFloat
  \captionsetup{labelformat=continued}
  \caption{\textbf{a,} Lower triangle: Jaccard overlap of measured genes; upper triangle: ESM2 similarity of perturbation sets, with diagonal entries equal to 1. A01 and A10 denote Adamson 10X001 and 10X010; D7 and D13 denote Dixit 7-day and 13-day; Rep denotes Replogle.
  \textbf{b,} Distributions of gene-wise mean expression differences ($\Delta$) between perturbed and matched control cells, plotted using shared histogram bins and density scales.
  \textbf{c,} Score differences for each alignment method relative to unaligned pooling (Direct) on the held-out Replogle test set (mean $\pm$ SD across six models and three seeds). Lower MSE indicates improvement; higher values indicate improvement for all other metrics.
  \textbf{d,} Score changes when ranking genes across pooled training cells versus averaging per-dataset ranks, evaluated on shared genes for held-out perturbations from A10 and Rep.
  \textbf{e,} Score changes after adding one training dataset evaluated on a common target test set, grouped by target test dataset (columns) and added source dataset (colors). Bars show the mean across seven starting subsets, three seeds, and six models; positive values indicate improvement (with MSE sign-flipped).
  \textbf{f,} Paired score changes grouped by model and test dataset across the four candidate source datasets (mean $\pm$ SD). Lower MSE indicates improvement.
  \textbf{g,} Prediction performance across 25\%, 50\%, 75\%, and 100\% retained source cells for Direct, Anchor, and MNN using the four source datasets and the Replogle test set.
  \textbf{h,} Post-hoc joint-distance analysis from the separate 1,805-gene experiments. Positive values indicate reduced distance between training and target data after combining datasets; squares denote composite gains ($n=34$) and circles denote losses ($n=31$). Response distances are computed against measured target responses, and the horizontal axis includes a break.}
\end{figure}

Control cells provided a reference for baseline gene expression without the targeted genetic perturbations in each study. Using control cells from the training studies and Replogle, Anchor (\hyperref[sec:control_alignment]{Methods}), mutual nearest neighbors (MNN)~\cite{haghverdi2018mnn} and scVI~\cite{lopez2018deep} estimated adjustments that were applied only to expression data from the training studies. Replogle's perturbed-cell expression data were reserved for evaluation and were not used to estimate these adjustments. Averaged across the six models and three seeds, mean direction accuracy increased from 57.4\% with Direct to 61.4--63.0\% across the three adjustment methods. All three also improved mean fold-change rank agreement and reduced mean response MSE (Fig.~\ref{fig:multisource}c).

We also examined which genes to retain for training when combining datasets with different gene coverage (Fig.~\ref{fig:multisource}a). We compared ranking variable genes after pooling balanced training cells with averaging rankings calculated separately for each dataset. Ranking genes after pooling slightly improved mean correlation and MSE on Adamson 10X010 and Replogle, but reduced top-50 DE recovery on Replogle (Fig.~\ref{fig:multisource}d). Thus, a gene selection strategy that improved average expression prediction did not necessarily improve identification of affected genes.

A separate question was whether including an additional dataset would improve prediction under a fixed preprocessing workflow. For each comparison, we trained each model with and without one additional dataset and evaluated both versions on the same held-out perturbation experiments. On Replogle, adding Adamson 10X010 improved all six displayed metrics on average, whereas adding Adamson 10X001 worsened all six (Fig.~\ref{fig:multisource}e). These means summarize paired changes across starting dataset combinations, six models and three seeds. The direction and size of the changes also varied across models (Fig.~\ref{fig:multisource}f). Thus, adding data from the same cell line was not sufficient to improve prediction in another study.

As adding a dataset changes both the number of cells and the mix of studies. To examine the effect of cell number within the same collection of studies, we varied the fraction of cells used from the same four datasets while evaluating the same Replogle test experiments (Fig.~\ref{fig:multisource}g). Increasing that fraction from 25\% to 100\% raised mean response correlation from 0.122 to 0.154 and reduced mean MSE from 0.054 to 0.042 under Anchor. These scores changed little under Direct or MNN. The benefit of using more cells therefore depended on how the data were processed in this comparison.

We also examined whether similarity to the target was associated with beneficial data additions. In a separate analysis, gains were more frequent when adding a dataset reduced both the difference between training and target perturbations, measured using ESM2 protein representations, and the difference between their expression responses (Fig.~\ref{fig:multisource}h). Together, these results support evaluating each proposed data combination with its preprocessing on an independent validation task.

% \FloatBarrier
% \phantomsection
\subsection{Evaluation metrics and training objectives}
\label{sec:metrics}

\begin{figure}[!htbp]
    \centering
    \includegraphics[width=.95\linewidth]{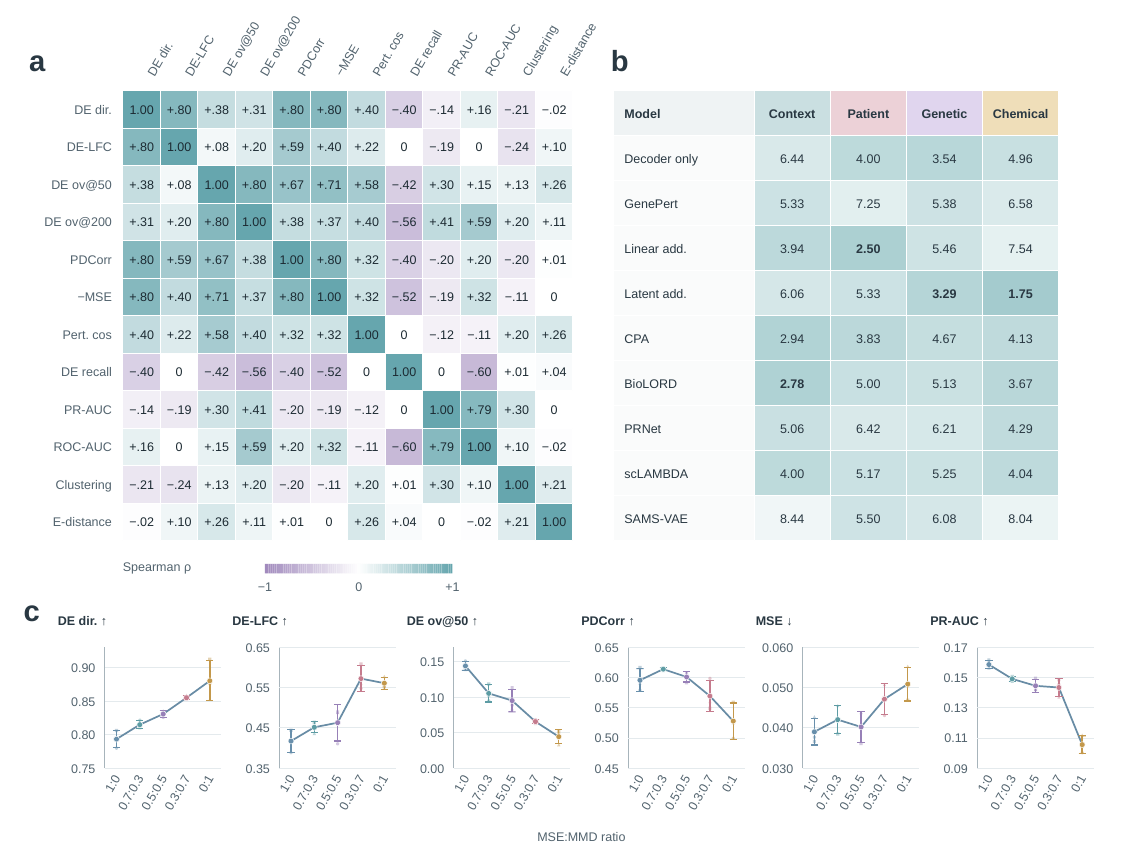}
    \caption{\textbf{Complementary metrics, task-specific ranks and a training-loss comparison.}
\textbf{a,} Median within-block Spearman correlations of model scores. Blocks are defined by experiment, setting and training seed. Negating MSE aligns its direction with the other scores in the correlation analysis.
\textbf{b,} Mean ranks across the six shared metrics DE direction, DE-LFC rank, DE ov@50, DE ov@200, PDCorr and response MSE. Raw scores are first averaged over three seeds within each setting, then nine models are ranked per metric. Ranks are averaged equally over metrics and OOD settings within each task; Random is excluded from these task means. Lower is better, with rank 1 best and MSE ranked in ascending order. Bold entries identify the lowest observed mean rank in each task.
\textbf{c,} PRNet on the fixed SciPlex3 A549 Random split at calibrated MSE:MMD ratios 1:0, 0.7:0.3, 0.5:0.5, 0.3:0.7 and 0:1. All settings select checkpoints by validation MSE. Points and error bars show the mean and sample SD over training seeds. MSE is shown on its raw scale, with lower values indicating smaller error.}
    \label{fig:metrics}
\end{figure}

In the preceding experiments, lower mean response MSE did not always coincide with higher top-50 DE overlap. To examine whether different metrics gave consistent assessments of model performance, we calculated Spearman correlations across models within each experiment, test setting and training seed, and summarized these correlations by their median (Fig.~\ref{fig:metrics}a). Response correlation and DE direction agreement generally favoured similar models, whereas fold-change rank agreement and top-50 DE overlap did not consistently favour the same models. We also compared task-level model ranks using the same six metrics, weighted equally. Under this scoring rule, Linear Additive had the best mean rank for held-out donors, whereas Latent Additive led prediction of unseen compounds; no model ranked first across all four evaluated task families (Fig.~\ref{fig:metrics}b).

We next asked whether changing the training objective could improve some aspects of prediction while worsening others, even when the model and task were unchanged. We compared five calibrated MSE--MMD weightings using PRNet on a fixed SciPlex3 A549 Random split, with all checkpoints selected by cell-level MSE on validation data. The MSE loss compared individual predicted and sampled observed cell-expression values, whereas MMD compared predicted and observed expression distributions within each perturbation condition. The response MSE used for evaluation instead compared predicted and observed mean expression changes relative to controls. Averaged over three training seeds, pure MMD increased DE direction accuracy from 79.3\% to 88.0\% relative to pure MSE, but reduced top-50 DE overlap from 14.4\% to 4.4\% (Fig.~\ref{fig:metrics}c). Fold-change rank agreement also improved, whereas PDCorr and PR-AUC decreased and response MSE increased.

Direction and fold-change rank agreement were evaluated on genes identified as differentially expressed in the observed data. Top-50 overlap instead measured agreement between the top-ranked DE gene lists obtained from predicted and observed data. On this fixed A549 Random split, replacing MSE with MMD improved mean direction and fold-change rank agreement on observed DE genes but reduced top-50 DE overlap.

%% file: Section/3_discussion.tex
% \FloatBarrier
% \phantomsection

\begin{figure}[!htbp]
    \centering
    \includegraphics[width=\linewidth]{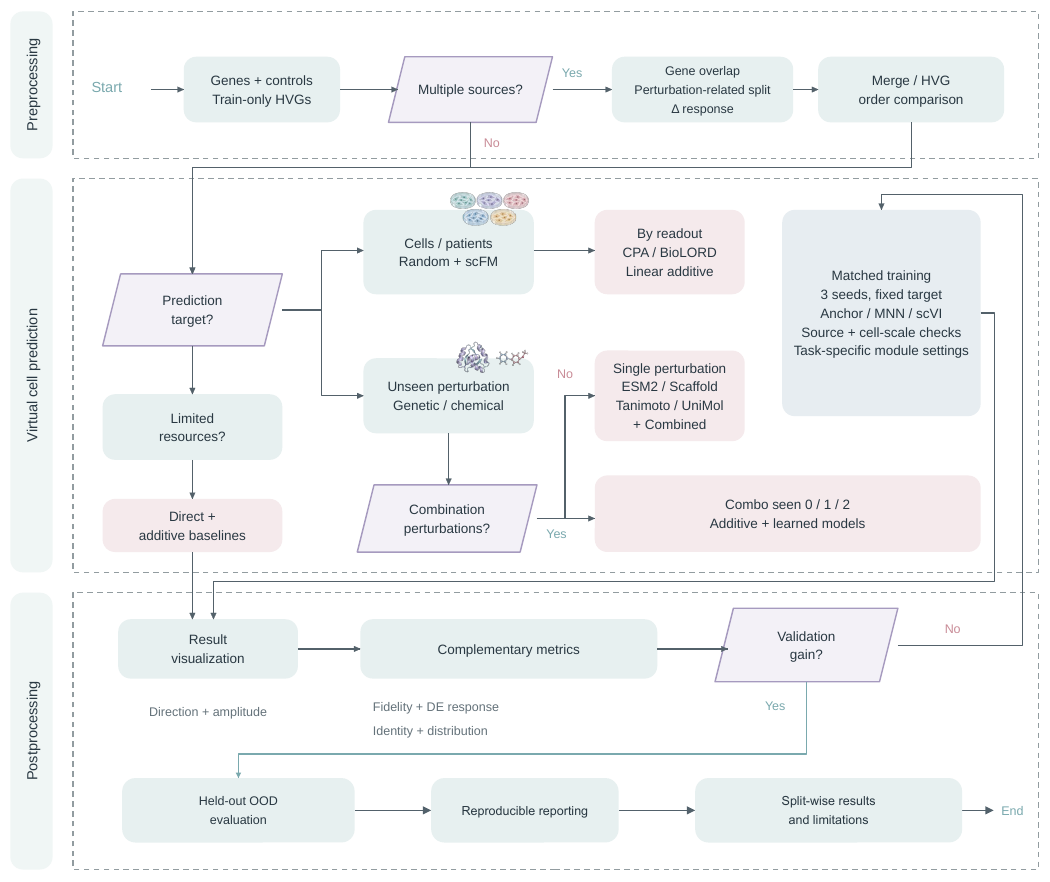}
    \caption{\textbf{Computational evaluation and reporting workflow.}
Define the prediction target and OOD split, harmonize genes and controls, and use training-only HVG selection. Compare preprocessing, training-source, and model choices on validation data with matched training budgets, three seeds, and fixed targets. Assess complementary response metrics, evaluate the selected configuration on the held-out OOD test, and produce a reproducible computational report.}
    \label{fig:guidance}
\end{figure}

\section{Discussion}
\label{sec:discussion}

Existing benchmarks show that model rankings vary across tasks and metrics~\cite{wei2025benchmarking,wu2025perturbench}. To move beyond these discrepancies, VCBench examines how model performance depends systematically on training-test similarity and multi-source data availability, leading to a structured workflow for experiment design and evaluation (Fig.~\ref{fig:guidance}). 

Taken together, our findings reveal key principles governing virtual cell generalization across biological contexts, interventions, and studies. Within individual systems, representation-guided holdouts demonstrate that models struggle more with distant cellular backgrounds and unseen perturbations, while simple linear baselines remain remarkably competitive by avoiding non-linear extrapolation failures. Across studies, multi-source integration depends less on raw cell-count scaling and more on whether external data can systematically narrow perturbation-embedding and response-distribution distances to the target. In addition, because these objectives and metrics pull in different directions, their selection must be carefully aligned with the intended biological application.

This study has several limitations that highlight important directions for future work. First, our evaluation focuses on a representative subset of models and data packages, and the main-text metrics were chosen to comprehensively capture descriptive trends rather than exhaust all possible criteria. Second, cross-study integration is currently centered primarily on K562 contexts. Although historical source-distance associations offer valuable descriptive insights, it remains an open challenge to use these metrics for a priori source selection before target responses are measured; validating these approaches across a wider array of diverse cell lines and independent cohorts will be essential to address this limitation. Finally, the current benchmark primarily targets steady-state transcriptional responses under specific splits, without systematically evaluating transfer across complex covariates such as varying temporal dynamics, drug dosages, or simultaneous background-intervention shifts.

Looking forward, advancing virtual cell modeling will require shifting from static endpoint prediction to capturing dynamic cellular state transitions and causal gene circuits. As broader time-series perturbation datasets and spatial multi-omic readouts become available, future frameworks can test whether transcriptional predictions faithfully translate to spatial tissue contexts and complex dynamic trajectories, ultimately bridging the gap toward reliable predictive models of cellular behavior.

%% file: Section/4_methods.tex
% \phantomsection
\section{Methods}
\label{sec:methods}

% \phantomsection
\subsection{Benchmark datasets}

We used harmonized data from scPerturb and the original studies~\cite{peidli2024scperturb}. Table~\ref{tab:datasets} lists the packages and the corresponding holdouts. We assign each held-out cell line, donor, genetic intervention or compound to a single partition. Cross-study comparisons reserve responses from a target study and use its control measurements as described below.

\begin{table}[htbp]
\centering\small
\caption{Data packages in the core benchmark. Counts refer to the indicated identity units.}
\label{tab:datasets}
\begin{tabularx}{\linewidth}{@{}lYY@{}}
\toprule
Package & Biological setting & Main analysis \\
\midrule
McFarland2020 & 170 cancer cell lines, three drugs & Cell-context OOD \\
Kang2018 CD4 T & Eight donors, IFNB stimulation & Cell-context OOD \\
Norman2019 & K562, single and paired gene perturbations & Perturbation-level OOD \\
SciPlex3 & 188 compounds, A549/K562/MCF7 & Perturbation-level OOD \\
A01 & Adamson 10X001, K562 & Dataset-level OOD \\
A10 & Adamson 10X010, K562 & Dataset-level OOD \\
D7 & Dixit 7-day, K562 & Dataset-level OOD \\
D13 & Dixit 13-day, K562 & Dataset-level OOD \\
Rep & Replogle K562 & Dataset-level OOD \\
\bottomrule
\end{tabularx}
\par\smallskip\footnotesize Core data sources include McFarland \cite{mcfarland2020multiplexed}, Kang \cite{kang2018multiplexed}, Norman \cite{norman2019exploring}, SciPlex3 \cite{srivatsan2020massively}, Adamson \cite{adamson2016multiplexed}, Dixit \cite{dixit2016perturb} and Replogle \cite{replogle2022mapping}.
\end{table}

\phantomsection
\subsection{Preprocessing and model training}

We use library-size normalization followed by $\log(1+x)$. The cell-context and single-perturbation experiments use prepared panels of 2,000 genes. For SciPlex3, we preserve the gene order of the reference HVG panel and normalize the retained expression to 10,000 counts per cell before the log transformation. The cross-study HVG comparison normalizes on the common gene set before projecting onto the selected genes.

For McFarland and Kang, we balance cell lines and donors through sample counts for controls and perturbed cells separately. The centroids for defining the splits use the original controls. We balance the number of treated cells per perturbation in Norman single-gene and per compound in SciPlex3, keeping controls unchanged in the upsampling step.

Within each task, models use the same expression features and perturbation representations. Table~\ref{tab:models} describes the implementations, including three baselines adapted from PerturBench~\cite{wu2025perturbench}. We trained each model with three random seeds and selected checkpoints by validation loss.

\begin{table}[htbp]
\centering\small
\caption{Model implementations used in VCBench.}
\label{tab:models}
\begin{tabularx}{\linewidth}{@{}lY@{}}
\toprule
Model & Benchmark component \\
\midrule
Linear Additive \cite{wu2025perturbench} & Adds a learned perturbation-dependent shift to control expression \\
Latent Additive \cite{wu2025perturbench} & Decodes the sum of encoded control and perturbation representations \\
Decoder Only \cite{wu2025perturbench} & Predicts expression from perturbation features \\
CPA \cite{lotfollahi2023predicting} & Combines latent effects additively before decoding.  \\
BioLORD \cite{haimovich2023biolord} & Decodes concatenated latent representations to predict expression \\
SAMS-VAE \cite{bereket2024modelling} & Adds an encoded perturbation effect to a sampled basal latent, bypassing the original sparsity mask \\
scLAMBDA \cite{jin2024sclambda} & Uses mutual information regularization to separate basal and perturbation representations \\
PRnet \cite{qi2024predicting} & Uses control expression and perturbation features to predict gene-wise expression distributions \\
GenePert (MLP variant) \cite{chen2024genepert} & Predicts expression from perturbation features with a two-layer MLP \\
\bottomrule
\end{tabularx}
\end{table}

\phantomsection
\subsection{Cell-context holdouts}

We compare random and guided holdouts of cell lines in McFarland and donors in Kang. We represent each background by the centroid of its control-cell embeddings and measure its cosine distance to the nearest training background. Models that use control expression receive target controls at inference. Perturbation responses from held-out backgrounds are reserved for validation or test.

For McFarland, we use STACK~\cite{dong2026stack}, scGPT~\cite{cui2024scgpt} and UCE~\cite{rosen2026uce} to construct guided splits that increase nearest-training distance while balancing the validation and test sets. For Kang, we use STACK embeddings of CD4 T cell controls to select held-out donor pairs, with one donor for validation and one for testing. The random and guided splits use the same partition sizes and predictor inputs.

\phantomsection
\subsection{Perturbation holdouts}

For Norman single-gene interventions, we compare random holdouts with splits that increase nearest-training cosine distance in ESM2 protein sequence space~\cite{lin2023evolutionary}. The splits use the same partition sizes and control assignments. For gene pairs, we reserve all interventions containing designated held-out genes. We also reserve some pairs whose individual genes occur in training. Test pairs are grouped by how many of their genes occur in training, either alone or in another pair. We measure nearest-training ESM2 distance for each component gene separately. Controls are partitioned independently of perturbed cells in both Norman tasks.

In SciPlex3, each compound is assigned to a single partition across doses, times and cell lines. All five designs use the same partition sizes and preserve coverage of the evaluated contexts. The Tanimoto and UniMol splits increase nearest-training distance using Morgan fingerprints~\cite{rogers2010extended} and UniMol embeddings~\cite{zhou2023unimol}, respectively. The Scaffold split keeps compounds with the same Bemis--Murcko scaffold together~\cite{bemis1996frameworks}. Combined keeps compounds with identical fingerprints or shared nonempty scaffolds together and increases both distances relative to Random.

\phantomsection
\subsection{Cross-study transfer in K562}

Within each comparison, we fix the target test set and include only controls from the target in training. We use Linear Additive, Latent Additive, CPA, BioLORD, PRNet and SAMS-VAE. We compare studies by the Jaccard overlap of measured genes, the mean pairwise cosine similarity between ESM2 embeddings of perturbed genes and the distribution of expression changes relative to matched controls.

\phantomsection\label{sec:control_alignment}
With Replogle as the target and A01, A10, D7 and D13 as sources, we compare direct concatenation with three alignment methods fitted on training controls. Anchor transforms the log-normalized expression $x_{s,g}$ of gene $g$ in each cell from source $s$ as
\begin{equation}
 x^{\mathrm{aligned}}_{s,g} = \mu_{t,g} +
 \frac{\sigma_{t,g}+\epsilon}{\sigma_{s,g}+\epsilon}(x_{s,g}-\mu_{s,g}),
 \qquad \epsilon=10^{-4}.
\end{equation}
Here $\mu$ and $\sigma$ are the gene-wise means and standard deviations of the training controls from source $s$ and target $t$. We apply the transform to every cell in each source. MNN~\cite{haghverdi2018mnn} estimates source corrections in PCA space, and scVI~\cite{lopez2018deep} estimates corrections by decoding under the observed and target batch labels. The methods correct source expression only, and the prediction models retain study identity.

For the HVG comparison, we rank genes within each dataset and average the ranks, or rank genes after pooling the same balanced sample of training cells. Both workflows train on the same cells and are scored on their shared genes. In the source-addition comparisons, we fix the HVG panel using target training controls and compare models trained with and without one candidate source on the same target test set. To examine source cell count separately, we vary the fraction of cells retained from the same four studies with Replogle as the target. 

\phantomsection
\subsection{Response evaluation}
\label{sec:metricdefinitions}

We use the bundled \texttt{cell\_eval} v0.6.6~\cite{adduri2026state}. Differential expression (DE) uses Wilcoxon tests at FDR $<0.05$. DE direction and fold-change rank agreement use genes significant in the observed data. The output catalog is in Supplementary Table~\ref{tab:metriccatalog}.

We calculate expression responses as mean changes relative to matched controls, separately for predicted and observed expression. Response MSE (\texttt{mse\_delta}) and PDCorr compare the two response vectors across genes by mean squared error and Pearson correlation, respectively. Expression MSE compares the expression means directly.

For the direction and amplitude comparison on SciPlex3 Combined, we restrict each condition to genes with significant DE and a nonzero observed response. Direction accuracy is the fraction with matching predicted and observed signs. Amplitude is the $L_2$ norm of the predicted response divided by the $L_2$ norm of the observed response over those genes. We weight all eligible condition--seed records equally.

\phantomsection
\subsection{PRNet loss weighting}
\label{sec:metriccomparison}

We vary the loss on a fixed Random split of SciPlex3 A549 at 10~$\mu$M, with matched dose-zero controls and 2,000 genes. The training objective is
\begin{equation}
 \mathcal{L}_{\alpha}=\alpha\,\mathrm{MSE}+(1-\alpha)\,c\,\mathrm{MMD}^{2},
 \qquad \alpha\in\{1,0.7,0.5,0.3,0\}.
\end{equation}
Here $c$ is the ratio of median MSE to median MMD$^2$ over up to 20 initial training batches for each seed and is shared across all weights. We compute MMD$^2$ within each perturbation condition and average across conditions, using three RBF kernels with bandwidths of 0.5, 1 and 2 times the training median squared distance per gene. Each batch uses 16 perturbation conditions with eight treated cells per condition and matched controls. Architecture, sampling and optimization stay fixed, and validation MSE determines early stopping and the selected checkpoint.

\phantomsection
\section{Data availability}

Source data accompany the manuscript. Harmonized expression data are available through scPerturb~\cite{peidli2024scperturb} at \url{https://zenodo.org/records/13350497} and from the original studies.

\phantomsection
\section{Code availability}

The VCBench framework is available at \url{https://github.com/maoxinjie/VCBench/}.

%% file: Section/X_appendix.tex
% \phantomsection
\section{Supplementary information}

\setcounter{table}{0}
\renewcommand{\thetable}{S\arabic{table}}
\renewcommand{\theHtable}{S\arabic{table}}

\phantomsection
\subsection{Evaluation metrics}

\begingroup\small
\begin{longtable}{@{}p{0.30\linewidth}p{0.52\linewidth}l@{}}
\caption{Evaluator outputs and score directions. DE counts are diagnostic.}\label{tab:metriccatalog}\\
\toprule
Display name & Source key & Direction\\
\midrule\endfirsthead
\multicolumn{3}{l}{\textbf{Table \thetable\ (continued)}}\\
\toprule Display name & Source key & Direction\\\midrule\endhead
\bottomrule\endfoot
DE dir. & \texttt{de\_direction\_match} & Higher \\
DE-LFC rank & \texttt{de\_spearman\_lfc\_sig} & Higher \\
DE-count correlation & \texttt{de\_spearman\_sig} & Higher \\
DE recall & \texttt{de\_sig\_genes\_recall} & Higher \\
DE PR-AUC & \texttt{pr\_auc} & Higher \\
DE ROC-AUC & \texttt{roc\_auc} & Higher \\
DE ov@50 & \texttt{overlap\_at\_50} & Higher \\
DE ov@100 & \texttt{overlap\_at\_100} & Higher \\
DE ov@200 & \texttt{overlap\_at\_200} & Higher \\
DE ov@500 & \texttt{overlap\_at\_500} & Higher \\
DE ov@N & \texttt{overlap\_at\_N} & Higher \\
DE prec@50 & \texttt{precision\_at\_50} & Higher \\
DE prec@100 & \texttt{precision\_at\_100} & Higher \\
DE prec@200 & \texttt{precision\_at\_200} & Higher \\
DE prec@500 & \texttt{precision\_at\_500} & Higher \\
DE prec@N & \texttt{precision\_at\_N} & Higher \\
MSE & \texttt{mse} & Lower \\
Response MSE & \texttt{mse\_delta} & Lower \\
MAE & \texttt{mae} & Lower \\
Response MAE & \texttt{mae\_delta} & Lower \\
PDCorr & \texttt{pearson\_delta} & Higher \\
Pert. cos & \texttt{discrimination\_score\_cosine} & Higher \\
Pert. L1 & \texttt{discrimination\_score\_l1} & Higher \\
Pert. L2 & \texttt{discrimination\_score\_l2} & Higher \\
E-distance correlation & \texttt{pearson\_edistance} & Higher \\
Clustering AMI & \texttt{clustering\_agreement} & Higher \\
True DE count & \texttt{de\_nsig\_counts\_real} & Diagnostic \\
Predicted DE count & \texttt{de\_nsig\_counts\_pred} & Diagnostic \\
\end{longtable}
\endgroup

\phantomsection
\subsection{Metric correlations and model features}
\newcounter{suppnote}\refstepcounter{suppnote}\label{supp:metricsarchitecture}

For each metric pair, we computed Spearman correlations across models within the same experiment, setting and seed. Table~\ref{tab:metriccorr} reports the median across groups with at least four models and variation in both scores.

We compared model rank percentiles by architectural feature. Matched controls and residual connections were associated with better ranks in cell-context tasks. Direct prediction from perturbation features was associated with better ranks in genetic tasks. Several features occur together, so these comparisons describe associations across models.

\begin{table}[htbp]
\centering\scriptsize
\setlength{\tabcolsep}{3pt}
\caption{Median Spearman correlations. Column numbers match row numbers. We reverse the sign of response MSE. E-distance correlates predicted and observed energy distances to controls.}
\label{tab:metriccorr}
\begin{tabular}{@{}lrrrrrrrrrrrr@{}}
\toprule
Metric & 1 & 2 & 3 & 4 & 5 & 6 & 7 & 8 & 9 & 10 & 11 & 12\\
\midrule
1. DE dir. & 1.00 & 0.80 & 0.38 & 0.31 & 0.80 & 0.80 & 0.40 & -0.40 & -0.14 & 0.16 & -0.21 & -0.02\\
2. DE-LFC & 0.80 & 1.00 & 0.08 & 0.20 & 0.59 & 0.40 & 0.22 & 0 & -0.19 & 0 & -0.24 & 0.10\\
3. DE ov@50 & 0.38 & 0.08 & 1.00 & 0.80 & 0.67 & 0.71 & 0.58 & -0.42 & 0.30 & 0.15 & 0.13 & 0.26\\
4. DE ov@200 & 0.31 & 0.20 & 0.80 & 1.00 & 0.38 & 0.37 & 0.40 & -0.56 & 0.41 & 0.59 & 0.20 & 0.11\\
5. PDCorr & 0.80 & 0.59 & 0.67 & 0.38 & 1.00 & 0.80 & 0.32 & -0.40 & -0.20 & 0.20 & -0.20 & 0.01\\
6. $-$MSE & 0.80 & 0.40 & 0.71 & 0.37 & 0.80 & 1.00 & 0.32 & -0.52 & -0.19 & 0.32 & -0.11 & 0\\
7. Pert. cos & 0.40 & 0.22 & 0.58 & 0.40 & 0.32 & 0.32 & 1.00 & 0 & -0.12 & -0.11 & 0.20 & 0.26\\
8. DE recall & -0.40 & 0 & -0.42 & -0.56 & -0.40 & -0.52 & 0 & 1.00 & 0 & -0.60 & 0.01 & 0.04\\
9. PR-AUC & -0.14 & -0.19 & 0.30 & 0.41 & -0.20 & -0.19 & -0.12 & 0 & 1.00 & 0.79 & 0.30 & 0\\
10. ROC-AUC & 0.16 & 0 & 0.15 & 0.59 & 0.20 & 0.32 & -0.11 & -0.60 & 0.79 & 1.00 & 0.10 & -0.02\\
11. Clustering & -0.21 & -0.24 & 0.13 & 0.20 & -0.20 & -0.11 & 0.20 & 0.01 & 0.30 & 0.10 & 1.00 & 0.21\\
12. E-distance & -0.02 & 0.10 & 0.26 & 0.11 & 0.01 & 0 & 0.26 & 0.04 & 0 & -0.02 & 0.21 & 1.00\\
\bottomrule
\end{tabular}
\end{table}

\newpage
\phantomsection
\subsection{Source addition}

\begin{table}[htbp]
\centering\scriptsize
\setlength{\tabcolsep}{3pt}
\caption{Mean changes after adding a study to the Direct baseline. Positive values indicate improvement, including lower response MSE.}
\label{tab:sourceaddition}
\begin{tabular}{@{}llrrrrrrr@{}}
\toprule
Test & Added & DE dir. & DE-LFC & ov@50 & ov@200 & PDCorr & MSE red. & Pert. cos\\
\midrule
A10 & A01 & +0.0030 & +0.0182 & +0.0019 & +0.0044 & +0.0033 & -0.0273 & +0.0065\\
A10 & D13 & +0.0090 & +0.0151 & +0.0003 & -0.0002 & +0.0102 & +0.0064 & -0.0018\\
A10 & D7 & -0.0015 & -0.0084 & +0.0015 & -0.0005 & -0.0024 & +0.0434 & +0.0060\\
A10 & Rep & -0.0032 & -0.0009 & +0.0113 & +0.0158 & +0.0170 & +0.1362 & +0.0324\\
Rep & A01 & -0.0073 & -0.0091 & -0.0018 & -0.0022 & -0.0065 & -0.0325 & -0.0013\\
Rep & A10 & +0.0431 & +0.0881 & +0.0171 & +0.0228 & +0.0441 & +0.0959 & +0.0197\\
Rep & D13 & +0.0013 & +0.0080 & -0.0004 & -0.0002 & +0.0031 & -0.0045 & -0.0018\\
Rep & D7 & +0.0040 & +0.0035 & +0.0002 & +0.0008 & +0.0009 & +0.0382 & +0.0036\\
\bottomrule
\end{tabular}
\end{table}